\documentclass[11pt,twoside]{article}
\usepackage{asp2010}

\resetcounters

\begin{document}

\title{Dust in AGB stars: transparent or opaque?}
\author{Sara Bladh$^1$, Susanne H\"ofner$^1$, and Bernhard Aringer$^2$
\affil{$^1$Department of Physics \& Astronomy, Uppsala University, Sweden}
\affil{$^2$Astronomical Observatory of Padova, INAF, Italy}}

\begin{abstract}
It is commonly believed that winds of cool giants in their late evolutionary stages are driven by radiative pressure on dust grains, but the actual grain species responsible for driving winds of M-type AGB stars are still a matter of debate (see H\"ofner, this volume). The optical properties of the dust particles responsible for initiating the outflow affect the stellar spectra in two ways: (i) indirectly, through their influence on the dynamical structure of the atmosphere/envelope and the resulting molecular features, and (ii) directly, by changes of the spectral energy distribution due to absorption and scattering on dust grains. The qualitative differences in the energy distributions of C-type and M-type AGB stars in the visual and near-infrared regions suggest that the dust particles in oxygen rich atmospheres are relatively transparent to radiation. By using detailed dynamical models of gas and radiation combined with a simple description for the dust opacity (which can be adjusted to mimic different wavelength dependencies and condensation temperatures) and also by adjusting the fraction of the opacity that is treated as true absorption, we investigate which dust properties produce synthetic photometry consistent with observations. The goal of this study is to narrow down the possible dust species that may be driving the winds in M-type AGB stars.
\end{abstract}
\section*{The parameterized dust opacity}
In the current study, the dust opacity which determines the radiation pressure is approximated by a wavelength dependent maximum opacity, $\kappa_{max}(\lambda)$, multiplied by a degree of condensation $f_c$ that changes with distance from the
star and time, i.e., 
$$\kappa_{dust}(\lambda)=\kappa_{max}(\lambda)\cdot f_c(r,t) = \kappa_{max}(\lambda)\cdot (1+\exp{(T_d(r,t)-T_c)/\delta T})^{-1}$$
The degree of condensation is determined by the grain temperature $T_d$ (given by radiative equilibrium in the time-dependent outflow), the assumed condensation temperature $T_c$ of the grain material and a variable $\delta T$ which adjusts how fast the condensation degree rises \citep[cf.][]{1988ApJ...329..299B}. A higher condensation temperature moves the dust-forming zone closer to the surface of the star and a larger value of $\delta T$ makes the dust forming zone wider. The dust opacity can be set to mimic different typical optical behaviors by giving $\kappa_{max}$ different wavelength dependencies, e.g., constant in the grey case or as a power law, $\kappa_{max}(\lambda)\propto\lambda^{-p}$.

Since the opacity relevant for radiation pressure is a combination of absorption and scattering terms, the fraction of the total opacity which is treated as true
absorption in the radiative transfer is also varied. If this fraction is set to zero the radiation pressure will still add momentum but there will be no direct influence on the energy distribution, simulating transparent dust grains. The scattering cross-section of the dust particles (which is negligible for small grains) will be significant for grain sizes comparable to the wavelength corresponding to the peak intensity of the stellar radiation.
\section*{Models and first results}
The results presented here were obtained using a state of the art dynamical code \citep[cf.][]{2008A&A...491L...1H} together with the parameterized dust opacity described above and the gas opacity generating code COMA \citep{2009A&A...503..913A}. We want to point out that this approach gives a fully consistent dynamical structure of the atmosphere and wind-forming region, including the dust-free molecular layers in the extended outer atmosphere. The purpose of using a parameterized dust description is to narrow down the range of possible dust species responsible for driving the outflow by studying the resulting SEDs and colors. A first comparison with observations shows that models with a noticable fraction of dust absorption have extremely high (V-K) colours, independent of $T_c$, whereas a model with scattering only (no absorption by dust grains) lies well within the range of observed colours for $T_c=1500$K. In addition to the models with a parameterized dust opacity, we also show the colours of a  consistent dynamical model with a wind driven by large Fe-free silicate grains \citep{2008A&A...491L...1H} which fits rather well with observations.

\begin{figure}
\begin{center}
 \includegraphics[width=89mm]{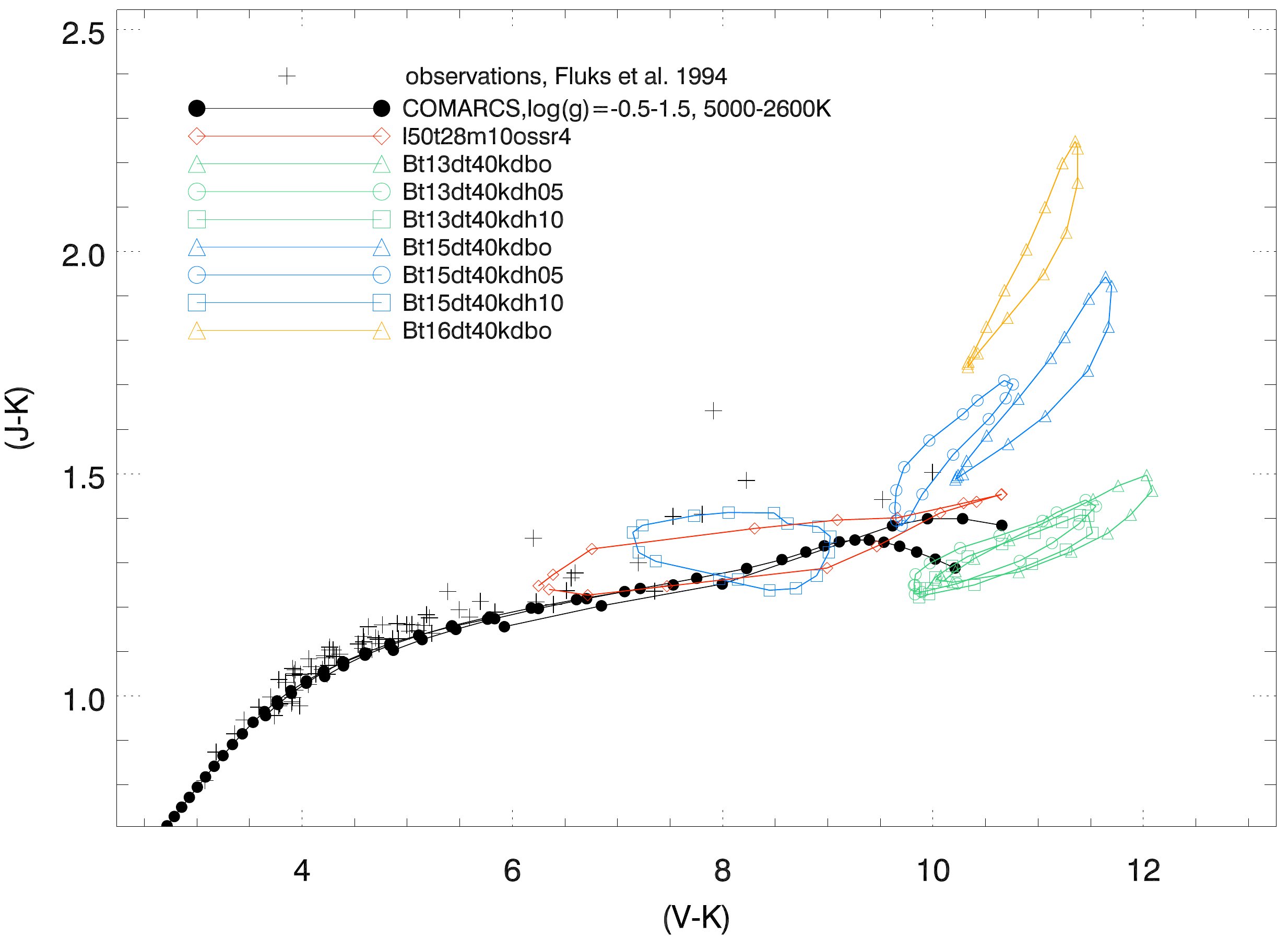}
\end{center}
\footnotesize
 {\bf Fig.} 
 The filled circles show the photometric values from hydrostatic models with different $T_{\rm{eff}}$ and log(g) hydrostatic M star models similar to the C models in Aringer et al. (2009). The data from the dynamical models with parametrized dust opacity are plotted for 100\% absorption (triangles), 50\% absorptions (circles) and 0\% absorption (squares) with $\kappa_{max} = 3\,$[cm$^2$/g] (grey). The parameter space includes 3 different condensation temperatures (1300K,1500K and 1600K) and $\delta T$ is set to 40K. All dynamical models are based on the same hydrostatic initial model with $T_{\ast} = 2800$K, $L_{\ast} = 5000 L_{\odot}$ and $M_{\ast} = 1 M_{\odot}$. Model l50t28m10ossr4 is a consistent dynamical model with a wind driven by large Fe-free silicate grains.
\end{figure}

\bibliography{bladh}

\begin{thebibliography}{}
\expandafter\ifx\csname natexlab\endcsname\relax\def\natexlab#1{#1}\fi
\expandafter\ifx\csname url\endcsname\relax
  \def\url#1{\texttt{#1}}\fi
\expandafter\ifx\csname urlprefix\endcsname\relax\def\urlprefix{URL }\fi
\providecommand{\eprint}[2][]{\url{#2}}

\bibitem[{{Aringer et al}(2009)}]{2009A&A...503..913A}
{Aringer et al}, B. 2009, \aap, 503, 913. \eprint{0905.4415}

\bibitem[{{Bowen}(1988)}]{1988ApJ...329..299B}
{Bowen}, G.~H. 1988, \apj, 329, 299

\bibitem[{{H{\"o}fner}(2008)}]{2008A&A...491L...1H}
{H{\"o}fner}, S. 2008, \aap, 491, L1

\end{thebibliography}

\end{document}